# Evidence From Gravitational Lensing For A Non-Thermal Pressure Support In The Cluster of Galaxies A2218

Abraham Loeb and Shude Mao
Harvard-Smithsonian Center for Astrophysics, 60 Garden Street, Cambridge, MA 02138

## ABSTRACT

The central mass distribution of clusters of galaxies can be inferred from gravitationally lensed arcs with known redshifts. For the Abell cluster 2218, this method yields a core mass which is larger by a factor of $2.5 \pm 0.5$ than the value deduced from X-ray observations, under the assumptions that the gas is supported by thermal pressure and that the cluster is spherical. We show that a non-thermal pressure support is the most plausible explanation for this discrepancy. Such a pressure can be naturally provided by strong turbulence and equipartition magnetic fields ($\sim 50\mu$G) that are tangled on small spatial scales ($\ll$ 10 kpc). The turbulent and magnetic pressures do not affect the measured Sunyaev-Zel'dovich effect for this cluster. Intracluster magnetic fields with a comparable magnitude ($\sim 10^{1-2}\mu$G) have already been detected by Faraday rotation in other clusters. If generic, a small-scale equipartition magnetic field should affect the structure of cooling flows and must be included in X-ray determinations of cluster masses.

## 1. Introduction

The central mass distribution in clusters of galaxies can be inferred in an unambiguous way from gravitational lensed arcs with known redshifts (Grossman & Narayan 1989). Surprisingly, this method yields a core radius which is smaller by a factor of a few than the value deduced from X-ray observations (e.g., Jones & Forman 1984; Edge & Stewart 1991) under the standard assumptions of thermal hydrostatic equilibrium and spherical symmetry. This discrepancy was highlighted recently by Miralda-Escudè & Babul (1994). Possible explanations to the discrepancy involve different violations of the standard assumptions, such as deviations from sphericity in the cluster mass distribution or non-thermal contributions to the gas pressure from magnetic fields and bulk motions. In this paper we show quantitatively that the most plausible explanation to the above discrepancy in the rich cluster A2218 is the existence of turbulence and equipartition magnetic fields in the cluster core.

Strong magnetic fields were detected in the past in many clusters of galaxies through Faraday rotation of radio sources (Kim et al. 1990, 1991; Taylor & Perley 1993). The detected fields have a typical magnitude of a few $\mu$G and a coherence length of $10^{1-2}$ kpc. Although significant, these fields are still an order of magnitude smaller than the equipartition values needed to make them dynamically important for the support of the cluster gas. However, the Faraday rotation method is limited from probing fields that are tangled on small spatial scales, because field reversals along the line of sight cancel out in the observed rotation measure. The possibility therefore remains that somewhat stronger magnetic fields, tangled on small scales ($\lesssim 10$ kpc), make a significant contribution to the pressure support of the cluster gas. Indeed, evidence for tangled magnetic fields with magnitudes as high as $\sim 10^{1-2}\mu$G was found in several clusters (Dreher et al. 1987; Perley & Taylor 1991; Taylor & Perley 1993; Ge & Owen 1993). Such fields should result in a discrepancy between the thermal pressure of the gas and the depth of the cluster potential well. In this work we argue that this discrepancy was in fact detected through gravitational lensing in the Abell cluster 2218.

The outline of this paper is as follows. In §2 we formulate the hydrostatic equilibrium equations for a hot gas with an isotropic (small-scale) magnetic stress, embedded in a cluster with a general ellipsoidal geometry. We then infer empirically the density distribution of the gas according to the observed profile of its bremsstrahlung surface brightness. At hydrostatic equilibrium, the fitted gas distribution can then be used to find the net gravitational potential of the cluster. Gravitationally lensed arcs probe the total cluster mass enclosed within a cylinder at some projected radius. This mass can be related to the observed surface brightness of the X-ray gas using Poisson's equation. This empirical approach (cf. Fabricant, Rybicki & Gorenstein 1984) has the attractive feature that it does not enforce any particular form of the total mass distribution; instead, the lensing mass is obtained directly from the observed X-ray surface brightness profile. The method is applied in §3 to the Abell cluster 2218 for which there are a number of additional observational data, including the velocity dispersion of the member galaxies (Le Borgne et al. 1992) and measurements of the Sunyaev-Zel'dovich (SZ) effect (Birkinshaw & Hughes 1994). We show that the mass of the projected cluster core has to be larger by a factor of $2.5\pm0.5$ than the value inferred from the standard X-ray analysis. We then demonstrate that this discrepancy cannot be explained by an elongation of the cluster along the line of sight. On the other hand, the discrepancy can naturally result from strong turbulence and equipartition magnetic fields in the cluster core. In §4 we summarize these conclusions and discuss ways to test them empirically.

## 2. Empirical Relation Between The Lensing Mass and X-Ray Data

The standard description of the hot gas in clusters of galaxies assumes hydrostatic support by thermal pressure and spherical symmetry. To consider deviations from these standard assumptions we express the total gas pressure as the sum of thermal and non-thermal components $p = p_t + p_{nt}$, and assume a general ellipsoidal geometry for the cluster gas. The thermal pressure of the gas at a temperature $T$ is given by $p_t = \rho_g kT/\mu m_p$, where $\rho_g$ is its mass density, $k$ is the Boltzmann constant, $m_p$ is the proton mass, and $\mu \approx 0.6$ is its mean atomic weight. The non-thermal pressure may contain three separate parts (which are comparable in magnitude for the interstellar medium of the Galaxy), namely: magnetic pressure $p_B$, turbulent pressure $p_{turb}$, and cosmic-ray pressure $p_{ray}$. A magnetic field that is tangled on scales small compared to the cluster core radius ($\sim$ 100 kpc) yields a diagonal magnetic stress tensor with a pressure[1] $p_B = \langle|\vec{B}|^2\rangle/8\pi$. Turbulent motions and bulk velocities add $p_{turb} = \frac{1}{3}\langle\rho_g v^2\rangle$, while cosmic-rays confined by the magnetic field provide their own kinetic pressure $p_{ray} \lesssim p_B$.

The hydrostatic equilibrium equation is

$$\frac{1}{\rho_g}\vec{\nabla} p = -\vec{\nabla}\Phi, \tag{1}$$

where $\Phi$ is the gravitational potential of the cluster. Since gravitational lensing probes mainly the cluster core we assume for simplicity that

$$p_{nt} = \alpha\ p_t, \quad \alpha = const; \tag{2}$$

and that the gas is isothermal. Equations (1) and (2) yield,

$$\rho_g \propto \exp\left\{-\frac{\mu m_p \Phi}{(1+\alpha)kT}\right\}. \tag{3}$$

The total mass density of the cluster can then be inferred from Poisson's equation,

$$\rho_{tot} = \frac{1}{4\pi G}\nabla^2\Phi = \frac{(1+\alpha)kT}{4\pi G\mu m_p}\nabla^2 \ln \rho_g^{-1}, \tag{4}$$

if the gas density distribution is known from X-ray observations.

---

[1] The magnetic pressure contains two components, arising from the magnetic energy density $u_{\rm mag} = (B^2/8\pi)$ and the kinetic energy density $u_{\rm kin} = u_{\rm mag}$ of Alfven waves (Dewar 1970). Other modes are Landau damped (Kulsrud 1994). The magnetic virial theorem (Shu 1992) yields the total pressure $p_B = \frac{2}{3}u_{\rm kin} + \frac{1}{3}u_{\rm mag} = B^2/8\pi$.

The X-ray surface brightness of the cluster gas is obtained by a line integral of its bremsstrahlung emissivity,

$$I \propto \int \rho_g^2 \, dz, \tag{5}$$

where we use cylindrical $(r, z)$ coordinates around the line of sight. Typically, the observed X-ray brightness contours are close to circular. We therefore assume axial symmetry around the line of sight, and adopt the simplest model to describe deviations of clusters from sphericity. In particular, the gas density is assumed to be a function of the ellipsoidal coordinate,

$$m^2 = \frac{r^2}{r_c^2} + \frac{z^2}{z_c^2}, \tag{6}$$

where $(r_c, z_c)$ is the ellipsoidal core size. For the simple case of a spherical gas distribution, $m = R/R_c$ is the radial coordinate and $R_c = r_c = z_c$ is the core radius. With $\rho_g = \rho_g(m)$ we can use Abell's equation (e.g., Binney & Tremaine 1987) to invert equation (5),

$$\rho_g^2 \propto \int_{mr_c}^{\infty} \left(-\frac{dI}{dr}\right) \frac{dr}{\sqrt{r^2/r_c^2 - m^2}}. \tag{7}$$

Typically, the X-ray surface brightness profile $I(r)$ is well fitted by a functional form (e.g. Birkinshaw, & Hughes 1994; Bahcall & Lubin 1994, and references therein)

$$I(r) \propto \left(1 + \frac{r^2}{r_c^2}\right)^{1/2-3\beta} \frac{z_c}{r_c} \tag{8}$$

with $\beta = 0.5 - 0.9$. Equation (7) then yields a gas distribution of the form,

$$\rho_g(m) \propto (1 + m^2)^{-3\beta/2}. \tag{9}$$

Gravitational lensing probes the total mass enclosed within a cylinder of a particular radius $r = b$ along the line of sight. To find this mass we calculate the volume integral,

$$M(b) = \int_V \rho_{tot} \, dV = -\frac{(1+\alpha)kT}{4\pi G\mu m_p} \int_S \vec{\nabla} \ln \rho_g \cdot d\vec{S} = -\frac{(1+\alpha)kTb}{G\mu m_p} \int_0^{\infty} dz \, \frac{\partial \ln \rho_g}{\partial r}\Big|_{r=b}, \tag{10}$$

where we have substituted equation (4), and used Gauss's theorem to convert the volume integral into a surface integral on an infinite cylinder. Note that the lensing mass $M(b)$ depends only on the shape of the gas density distribution and not on its absolute normalization. Using equation (9) we finally obtain

$$M(b) = \frac{\pi kT}{G\mu m_p} \, \frac{3\beta}{2} \, \frac{b^2}{\sqrt{b^2 + r_c^2}} \, (1+\alpha) \, \frac{z_c}{r_c}. \tag{11}$$

This is our basic result. Observational measurements of $T, \beta, r_c, b$ and $M(b)$ can be used to determine the quantity $(1+\alpha)z_c/r_c \equiv \eta$. Values of $\eta \neq 1$ quantify the degree to which the standard assumptions concerning the cluster gas are violated.

Finally, we consider another probe of the thermal state of the cluster gas, namely the Sunyaev-Zel'dovich (SZ) effect. This effect results from Thomson scattering of the cosmic background photons by hot electrons in the optically-thin cluster. If the cluster has no peculiar motion, the single-scattering Doppler shift averages to zero to linear order in the electron velocity, since the distribution of electron velocities is isotropic at thermal equilibrium. However, the second-order Doppler effect does not vanish and is proportional to the square of the thermal velocity of the electrons, i.e. to $kT$. The scattering probability is proportional to the integral of the electron density along the line of sight, and thus the net SZ effect is linear in the electron pressure. The scattering of microwave photons to higher energies leads to a decrease in the Rayleigh-Jeans temperature of the cosmic background radiation $T_{RJ}$, $\Delta T_{RJ}/T_{RJ} \propto -\int p_t dz$. Subsonic turbulent velocities provide $v_{turb}^2 \lesssim (kT/m_p)$ and result in an effect that is smaller than the thermal effect of the electrons by the electron to proton mass ratio. The existence of magnetic fields and cosmic-rays can also be ignored in calculating the SZ distortion. Thus, for the density profile of equation (9) the distortion is given by,

$$\Delta T_{RJ}(r) \propto \left(1+\frac{r^2}{r_c^2}\right)^{1/2-3\beta/2} \frac{z_c}{r_c}, \tag{12}$$

with no reference to the non-thermal pressure of the gas.

Note that according to equations (8) and (12), the elongation of clusters does not change the radial shapes of either the X-ray surface brightness or the SZ distortion. The elongation only enhances both effects by the axis ratio factor $z_c/r_c$.

## 3. Application to A2218

Next we apply the results from §2 to study the conditions of the gas in the rich cluster A2218, for which there is a wealth of data, including the X-ray surface brightness profile of the cluster, the positions and redshifts of gravitationally lensed arcs, the Sunyaev-Zel'dovich effect, and the velocity dispersion of the member galaxies. We denote the Hubble constant by $H_0 = 100\ h$ km s$^{-1}$ Mpc$^{-1}$.

Abell 2218 is a cluster at a redshift $z = 0.175$ (Le Borgne et al. 1992) with a richness class 4, an X-ray luminosity in the 2-10keV band of $10^{45}$ erg s$^{-1}$ (David et al. 1993), a central proton density of $(5.4 \pm 0.5) \times 10^{-3} (h/0.5)^{1/2} (z_c/r_c)^{-1/2}$cm$^{-3}$ (Birkinshaw & Hughes 1994), and an X-ray temperature of $kT = 6.7^{+0.5}_{-0.4}$ keV (McHardy et al. 1990). At the redshift of the cluster, $1'' \equiv 1.9\ h^{-1}$ kpc. The optical image of this cluster reveals several gravitationally lensed arcs. One of the arcs is at a redshift of $z_s = 0.702$ with a critical angular radius of $\theta_{crit} = 20.8''$ centered at the X-ray peak of the cluster (Pellò et al. 1992; see also Miralda-Edcudé & Babul 1994). The SZ effect for A2218 was most recently analyzed by Birkinshaw & Hughes (1994), who improved previous work by McHardy et al. (1990) and derived a Hubble constant of $h = (0.65 \pm 0.25) \times (r_c/z_c)$. Birkinshaw & Hughes (1994) fitted the X-ray surface brightness profile using equation (8) and obtained the best fit with $\beta = 0.65 \pm 0.04$, and an angular core radius of $\theta_c = 60'' \pm 10''$. With these values, equation (8) also provides an excellent fit to the surface brightness profile for this cluster as recently measured by ROSAT (Stewart et al. 1993). We therefore adopt these values in the analysis that follows.

The observed arcs primarily constrain the total mass inside the cylinder of the critical radius $b$. The constraint is relatively insensitive to the mass distribution profile (e.g., Kochanek 1991), and can be expressed as

$$M(b) = \frac{c^2\theta_{crit}^2}{4G}\frac{D_L D_S}{D_{LS}}, \tag{13}$$

where $D_L, D_S$, and $D_{LS}$ are the angular diameter distances to the lens, to the source, and from the lens to the source, respectively (see Kochanek 1992 for the redshift dependence of $D$ on different cosmologies). Note that $b \equiv \theta_{crit} D_L$ and $r_c \equiv \theta_c D_L$. For A2218, equation (13) gives $M(b) = 0.32h^{-1} \times 10^{14} M_\odot$ assuming a cosmological density parameter $\Omega = 1$. This result has a negligible dependence on the underlying cosmology. In fact, for any cosmology with $0 \leq \Omega \leq 1$ and no cosmological constant ($\lambda = 0$), or for any flat cosmology with $\lambda \neq 0$, this mass constraint varies by less than 3%. Equation (13) assumes an axially-symmetric lens geometry. In reality, the contribution from the presence of a central cD galaxy and substructure in A2218, and the ellipticity needed to break the cylindrical symmetry and form arcs instead of rings, lower this mass estimate by $\approx 15\%$ (Miralda-Escudé 1994).

However, from equation (11) we obtain $M(b) = (0.11 \pm 0.02)(kT/6.7\mathrm{keV})\eta\ h^{-1} \times 10^{14} M_\odot$. In order for these two mass estimates to be consistent, we find

$$\eta \equiv (1+\alpha)\frac{z_c}{r_c} = 2.5 \pm 0.5, \tag{14}$$

where the error bars are dominated by the uncertainties in the X-ray observations and can be reduced considerably in the future.

The first possible origin for this value of $\eta$ is that the cluster is prolate. If the contribution of non-thermal pressure is negligible, then the axis ratio for the gas distribution must be $z_c/r_c \approx 2.5$. We then find, using equations (4) and (9), that the total mass distribution has an axis ratio of roughly 6:1 in the cluster core. The axis ratio for the total mass is larger than $z_c/r_c$ because the gas follows the potential which tends to be smoother than the underlying mass distribution (Binney & Tremaine 1987). Furthermore, the total mass distribution has an unphysical dumbbell shape at large radii for any value of $(z_c/r_c) > \sqrt{3/2}$, as long as the gas distribution is described by equation (9) (cf. Kassiola & Kovner 1993). The above properties of the total mass distribution do not seem to be plausible. With an axis ratio 6:1, the probability for a perfect alignment of the cluster along the line of sight (as required by the circular X-ray brightness contours) is only a few percent times the relatively small fraction of prolate clusters seen from the side (Jones & Forman 1992). In addition, such a configuration is unstable against bending modes (Merritt & Hernquist 1991). Therefore, a full account for the discrepancy by the prolateness of the cluster is unlikely.

The alternative explanation involves a non-thermal pressure support of the gas. This explanation is consistent with other observational data. The observed velocity dispersion of galaxies (Le Borgne et al. 1992) $1370^{+160}_{-120}\mathrm{kms}^{-1}$ is larger than $(kT/\mu m_p)^{1/2} = 1050^{+40}_{-30}\mathrm{kms}^{-1}$. In addition, the most recent determination of the Hubble constant from the SZ effect in A2118 (Birkinshaw & Hughes 1994) did not yield a value smaller by a factor of 2.5 from the range $0.5 \lesssim h \lesssim 1$, as would be expected according to equation (12) if the gas distribution was elongated along the line of sight rather than being supported by a non-thermal pressure.

In principle, all the potential sources of non-thermal pressure are equally viable in accounting for the lensing mass discrepancy. However, large bulk velocities can be excluded by other considerations. Supersonic rotational velocities would have flattened the X-ray brightness contours, while large radial velocities would have relaxed to equilibrium within a dynamical time ($\ll 10^{10}$yr). The existence of a strong turbulent pressure $p_{turb} \sim p_t$ is likely to be accompanied by equipartition magnetic fields, in analogy with the conditions in the interstellar medium of the Galaxy. We therefore conclude that equation (14) implies the existence of dynamically important turbulence and magnetic fields in A2218.

For example, $\alpha \approx 2$ may correspond to $p_B \approx p_{turb} \approx p_t$. This amounts to an equipartition magnetic field strength

$$B = 53\mu\mathrm{G}\left(\frac{n_e}{5\times10^{-3}\mathrm{cm}^{-3}}\right)^{1/2}\left(\frac{kT}{7\mathrm{keV}}\right)^{1/2}, \qquad (15)$$

where $n_e$ is the electron density. If this field is tangled on relatively small scales ($\lesssim 10$kpc), it would partially avoid detection by Faraday rotation measurements. Previous observations of other clusters (Dreher et al. 1987; Kim et al. 1990, 1991; Perley & Taylor 1991; Taylor & Perley 1993; Ge & Owen 1993) found evidence for fields that are somewhat weaker or comparable to the equipartition value ($10^{0-2}\mu$G) on large scales ($10^{0.5-2}$ kpc). Even relatively weak fields should eventually approach equipartition values at some radius as they are dragged inwards by the cooling flow of the cluster (Soker & Sarazin 1990).

The existence of strong fields results in synchrotron cooling of the cluster electrons. However, since the plasma frequency of the cluster, $f_p = 0.64$ kHz $(n_e/5 \times 10^{-3}\mathrm{cm}^{-3})^{1/2}$, is larger than the cyclotron frequency, $f_c = 0.14$ kHz $(B/50\mu\mathrm{G})$, the synchrotron emission is suppressed. The plasma cutoff is effective across the entire region where the field is in equipartition with the thermal pressure of the gas, because the ratio between the above frequencies scales as $\rho^{1/2}/B \propto (p_t/p_B)^{1/2}$. The synchrotron cooling time is therefore much longer than the bremsstrahulg cooling time in the cluster core, and any residual emission would occur at the unobservable frequency regime $\sim$ kHz. On the other hand, the synchrotron emission by cosmic rays is observable at microwave frequencies. From the fact that the measurement of the SZ effect (Birkinshaw & Hughes 1994) was not dominated by a synchrotron signal we conclude that $p_{ray} \ll p_t$. This leaves only magnetic fields and turbulence as viable sources of non-thermal pressure in A2218.

## 4. Conclusions

The core mass of the cluster of galaxies A2218 can be probed by two independent methods: gravitational lensing and X-ray observations. Assuming that the cluster is spherical and that the hot gas is supported by a thermal pressure, these two methods provide values that are different by a factor of $2.5 \pm 0.5$. This conclusion is in agreement with a recent study by Miralda-Escudè & Babul (1993). The result is not sensitive to the underlying cosmology. The main uncertainty is associated with the temperature at the cluster core, which was not resolved by early X-ray spectroscopy observations (McHardy et al. 1990). However, a recent observation by ROSAT shows a flat temperature profile, consistent with the modeling assumption of an isothermal cluster (Stewart & Edge 1993). The above mass discrepancy can be resolved by relaxing either of the two underlying assumptions about A2218. In §3 we showed that in order to explain the discrepancy by a deviation of the cluster from sphericity, its mass distribution must be elongated along the line of sight with an axis ratio of 6:1. Such a distribution is highly improbable, and also gravitationally unstable (Merritt & Hernquist 1991).

The alternative explanation involves a non-thermal pressure support of the gas. Large rotational velocities are excluded based on the circular X-ray brightness contours. Supersonic radial velocities would have been damped out within a dynamical time $\sim 10^{8-9}$yr. Indeed, the X-ray map of the cluster core does not show large scale inhomogeneities of the gas. In addition, cooling flow velocities are expected to be highly subsonic at the core radius.

However, there are other more plausible sources of non-thermal pressure. Turbulence and magnetic fields in equipartition with the thermal energy of the gas can fully account for the above discrepancy in a spherical cluster. This explanation is consistent with other observational data on A2218. In difference from a cluster elongation, the turbulent and magnetic pressures do not affect the Sunyaev-Zel'dovich effect, in line with the observations (Birkinshaw & Hughes 1994). The square of the observed velocity dispersion of galaxies (Le Borgne et al. 1992) is larger than $(kT/\mu m_p)$ by a factor of $1.7^{+0.4}_{-0.3}$; a value that may also reflect the different bias and virialization history of the galaxies relative to the gas in the cluster.

Magnetic fields with an amplitude $\gtrsim \mu$G were detected by Faraday rotation in many other clusters (Dreher et al. 1987; Kim et al. 1990, 1991; Owen et al. 1990; Perley & Taylor 1991; Taylor & Perley 1993; Ge & Owen 1993). In the Hydra A cluster, Taylor & Perley (1993) found a magnetic field $\sim 6\mu$G on a scale $\sim$ 100 kpc, and a tangled field as high as $\sim 30\mu$G on smaller scales. In the cluster A1795, Ge & Owen (1993) observed rotation measures exceeding 3000 rad m$^{-2}$, which translate to a field amplitude $> 20\mu$G. Similar results were obtained by Dreher et al. (1987) for the Cygnus A cluster, and by Perley & Taylor (1991) for the 3C295 cluster. The tangled magnetic field we predict for A2218 is comparable in amplitude to the inferred fields in these cases. Although the average rotation measure (RM) of such a tangled field is close to zero, the dispersion in the rotation measure should be large, $\langle \mathrm{RM}^2\rangle^{1/2} = 2140(n_e/5\times10^{-3})(B/50\mu\mathrm{G})(\ell/10\mathrm{kpc})(N/10)^{1/2}$ rad m$^{-2}$, where $N$ is the number of cells of size $\ell$ and field strength $B$ along the line of sight.

Aside from the existence of strong magnetic fields, our model predicts a considerable "$\beta-$discrepancy" (cf. Bahcall & Lubin 1994, and references therein) for A2218. The radial distributions of the gas and the galaxies should not be consistent with their relative velocity dispersions due to the non-thermal support of the gas.

Although galaxy-driven turbulence cannot readily produce the observed $\mu$G magnetic field (De Young 1992), there are a variety of field generating mechanisms operating in the cores of clusters (Taylor & Perley 1993). A natural source for the needed fields is dynamo amplification of seed galactic fields (Ruzmaikin et al. 1989; Kulsrud & Anderson 1992) due to the strong turbulence and shearing present during the collapse and virialization stages of the cluster. It is interesting to note that in an $\Omega = 1$ universe, the accretion of clumps of matter by the cluster continues at all times (Richstone, Loeb & Turner 1992) and can persistently excite turbulence in it. In analogy with the interstellar gas, the amplification of the magnetic fields would saturate at equipartition with the turbulent pressure. The resulting fields would then affect the structure of the cooling flow in the cluster (Soker & Sarazin 1990).

If future observations show that turbulence and magnetic fields are dynamically important not only in the inner parts but also in the outer parts of rich clusters, this would have a variety of interesting implications. First, it would increase the cluster mass to light ratio by a factor of a few and shift the resulting estimate of $\Omega$ from the range 0.2-0.3 (Binney & Tremaine 1987) to about unity. The baryonic mass fraction of clusters would consistently decrease to values $\lesssim 0.1$. Both of these changes would tend to make the cluster data agree with the predictions of inflation and standard nucleosynthesis. Second, an increase in the high mass tail of the cluster mass function would affect the comparison between observational data and popular cosmological models (Bahcall & Cen 1993). Finally, this would raise the possibility that magnetic fields have a non-negligible influence on structure formation in the universe.

We thank Daniel Eisenstein, Alyssa Goodman, Russell Kulsrud, Jordi Miralda-Escudé, Richard Mushotzky, and Ramesh Narayan for useful discussions.